# Architected kirigami metamorphosis

Yanbin Li,[1] Jie Yin[1,*]

[1]Department of Mechanical and Aerospace Engineering, North Carolina State University,

Raleigh, North Carolina 27695, USA

*Corresponding author. Email: jyin8@ncsu.edu

**One Sentence Summary:** kirigami-inspired evolutionary materials enable living-matter-like metamorphosis via kinematic bifurcation mechanism

**Abstract:** Kirigami, art of paper cutting, enables two-dimensional sheets transforming into unique shapes which are also hard to reshape once with prescribed cutting patterns. Rare kirigami designs manipulate cuts on three-dimensional objects to compose periodic structures with programmability and/or re-programmability. Here, we propose a new class of three-dimensional modular kirigami by introducing cuts on cuboid-shaped objects, based on which constructing two quasi-three-dimensional architected kirigamis with even-flat structural form. We demonstrate the proposed architected kirigamis are with rich mobilities triggered by kinematic bifurcations inherited from their composed modular kirigami, and can undergo living-matter-like metamorphosis evolving into miscellaneous transformable three-dimensional architectures and even a pluripotent platform capable of being re-programmed into curvature different surfaces through inverse design. Such metamorphic structures could find broad applications in reconfigurable metamaterials, transformable robots and architectures.

Metamorphosis is a biological evolution process of morphological changes in form and structure during the growth of some living systems such as animals, plants, and micro-organisms. Some organisms undergo incomplete metamorphosis with slight change in form, while others experience complete metamorphosis with abrupt shape transformation during their developmental stages (e.g., from a larva to a butterfly) to adapt to different living environments (1). Mathematically, the development of complete metamorphosis could be analogously considered as bifurcation that occurs in uniqueness-broken engineering systems, where it is often accompanied by abrupt topological transformation when independent variants reach critical values (2, 3). For example, Euler-elastica of straight rods bifurcates with deformed morphological branches (4); kinematic mechanisms manifest alternative transition paths accompanied by incremented mobilities at bifurcated thresholds (5, 6). Similar to the biological metamorphosis, some materials and structures exhibiting different levels of shape changing are proposed by means of folding (7), cutting (8, 9), rigid mechanisms (10), and buckling (11), i.e., they can change their shapes from an initially simple structure in two dimensions (2D) or three dimensions (3D) to sophisticated 3D forms imparted with new properties and functionalities (12) and broadly applied in reconfigurable metamaterials (13), self-reconfigurable robots (14, 15), morphing aerostructures (16) and four-dimensional (4D) printing (17, 18).

Kirigami, the ancient art of paper cutting, provides a powerful way to design metamorphic structures (19-20). Starting from a thin flat sheet with prescribed cutting patterns, it can rigidly reconfigure into multiple 2D/3D target shapes (21-22). Reversely, inverse design of cutting patterns can be prescribed to fit target topologies with various intrinsic curvatures (9, 23-25). However, once transformed, most kirigami structures are challenging to further reconfigure or reprogram into other morphologies due to limited internal degree of freedoms (DOF) and prescribed cutting patterns (21, 24-28). Similar with 2D flat kirigami thin sheet, the already proposed 3D kirigami (29) or analogous structure (30) face same transfiguration issue, i.e. limited structural reconfigurability except for volume expansion. Recent studies demonstrate that space-filling tessellation of 3D polyhedral origami modules creates 3D architected materials that can reconfigure into multiple 3D structural modes (37-38). However, such 3D prismatic materials exhibit limited mobilities and undergo incomplete structural metamorphosis that generates configurations without significant topology variation (33-35). Thus, it provokes the question whether one can design and create architected structures, concerning not only about enhancing structural mobilities by releasing internal structural constraints but also referring to the biological-like complete metamorphosis with bifurcation to achieve evolutionary structural variations and generate remarkably disparate structural morphologies.

Here, rather than morphing from 2D kirigami sheets (19-28) or tessellated 3D structures (29-35) to architectures with limited reconfigurability, we propose a new shape morphing strategy transforming from even-flat quasi-3D architected kirigami structures with rich mobilities to versatile 3D evolutionary architectures and targeted shapes via metamorphic kinematic mechanisms. As illustrated in Fig. 1 and Fig. S1A, we, firstly by introducing different number of 2D cutting planes directly on 3D cuboid-shaped objects, systemically explore the design of 3D modular kirigami formed with 4, 6 and 8 identical cuboid components. After a simple demonstration (38), the special 3D modular kirigami with kinematic bifurcations shown as Fig. 1B-1E is selected as basic unit to compose two periodic architected kirigami structures with metamorphic transfiguration features. Thus, in analogy to the development from cells to tissues and further to 3D architected organs, the selected 3D modular kirigami unit commences with a cube to "grow" into a cuboidal unit cell connected through a looped-linkage mechanism that allows reconfiguration and bifurcation into multiple 3D shapes (Fig. 1B to 1D). Assembly of the unit cells through unique planar tiling forms quasi-3D epithelial tissue-like structures with 3D metamorphic mechanism (36). Consequently, through both incomplete (non-bifurcated) and complete (bifurcated) metamorphosis, such originated planar structures are capable of evolving into varieties of 3D transformable architectures, part of which can be a pluripotent platform for further reprogramming and re-evolving into architected curved surface topologies through both forward and inverse design methods (Fig. S1D).

The selected 3D modular kirigami is formed by introducing four cutting planes and thus consists of eight rigid cubes connected through eight elastic torsional hinges as folding lines (Fig. 1B, top). The eight rotatable (8R) cubes can be simplified as linkages, and the spatial connections between links represent folding patterns. The combined linkage-hinge system forms a looped rigid mechanism, allowing rotation (folding) of rigid cubes around the hinges to transform spatially in 3D. The base unit demonstrates two kinematic mobilities according to the Grubler's Formula (37), thus enabling three divergent kinematic transition modes in terms of different number of rotated hinge pair(s): Mode 1 with 1R/2R linkage (Fig. 1B and Fig. S1B); Mode 2 with 6R linkage (Fig. 1c and Fig. S1C); and Mode 3 with 8R linkage mechanism (Fig. 1D and Fig. S1C). Mode 1 and Mode 2 are in analogy to the incomplete metamorphosis without kinematic bifurcation, while Mode 3 undergoes complete metamorphosis with multiple kinematic bifurcated paths as discussed later. To experimentally explore their transfiguration features, we 3D-print the prototypes of the rigid cube unit (Convex Objet-260) and hinge them by ultra-flexible plastic tapes to imitate free-rotational hinges (38).

Specially, Mode 3 is structurally indeterminate with two DOFs. After imposing structural symmetry constraint of the counterclockwise-ordered dihedral angles $\gamma_{ij}$ between two neighboring cube $i$ and cube in Fig. 1D-i, i.e., $\gamma_{23} = \gamma_{45} = \gamma_{67} = \gamma_{81}$, it becomes a single DOF mechanism, where the angle relations can be determined as $\gamma_{12} = \gamma_{56} = \gamma_{34} = \gamma_{78}$ ($0 \leq \gamma_{12} \leq 180°$) and $\gamma_{23}$ ($= \gamma_{45} = \gamma_{67} = \gamma_{81}) = \sin^{-1} [(1-\cos \gamma_{12})(1+\cos \gamma_{12})^{-1}]$ as a function of sole variable dihedral angle $\gamma_{12}$ (38).

Interestingly, for the single DOF mechanism in Mode 3, multiple sequential kinematic bifurcations occur during the transition process from state i to ix as shown in Fig. 1D, which transform from looped 8R linkage (Phase 1, left column in Fig. 1D) to looped 4R linkage (Phase 2, middle column in Fig. 1D), and further to non-looped 2R linkage mechanism (Phase 3, right column in Fig. 1D) at multiple bifurcation points. Notably, state ii with $\gamma_{12} = 90°$ is a bifurcation point that can bifurcate into two branched kinematic paths, i.e., from state ii to iv or to v as shown in Phase 2, where we observe retained dihedral angles of $\gamma_{23}$ ($= \gamma_{45} = \gamma_{67} = \gamma_{81}) = 90°$, consequently, its revolving joint number decreases from 8R to 4R. Furthermore, as the pore is closed, it encounters another bifurcation point at state vi with $\gamma_{12}$ or $\gamma_{34} = 180°$, which is an indeterminate classical 2R line-linkage mechanism (37) with two independent mobilities, thus engendering three combinatoric kinematic paths in Phase 3 without an internal loop, i.e., from state vi to vii or to viii or to ix shown in the right column of Fig. 1D. Quantitatively, the dihedral angle relations of the base unit during transition Mode 3 can be analytically determined based on the closed loop mechanism of $\sum_{i=1}^{8} \mathbf{T}_{ij} = \mathbf{I}$ (33) with the results shown in Fig. 1E, where $\mathbf{I}$ is a $4 \times 4$ identity matrix and $\mathbf{T}_{ij}$ is a 4×4 homogeneous transformation matrix determining the relative positions between cubes $i$ and cube $j$, see details in (38) and Fig. S2.

Theoretically, the kinematic features of the base unit including transformation and bifurcation can be exactly captured and predicted by the transformation matrix $\mathbf{T}_{ij}$ from one state to another with its components only dependent on the dihedral angles $\gamma_{ij}$ (38), where $\mathbf{T}_{ij}$ can be decomposed into the product of decisive rotations $\mathbf{T}_\gamma$ and constant translations $\mathbf{T_L}$ considering its sole rotational kinematics, i.e., $\mathbf{T}_{ij} = \mathbf{T}_\gamma \mathbf{T}_L$. To capture the bifurcation points, we introduce a numerical perturbation on the dihedral angle $\gamma_{ij}$ with an infinitesimal increment $\Delta\gamma_{ij}$ to $\mathbf{T}_\gamma$, which generates a coefficient matrix $\mathbf{M}$ in terms of the closed loop mechanism. Based on the single value decomposition method (34), $\mathbf{M}$ can be further decomposed into $\mathbf{M} = \mathbf{UVW}^T$, where $\mathbf{U}$ and $\mathbf{W}$ are square orthogonal matrices containing the left- and right-singular vectors, while $\mathbf{V}$ is a rectangular matrix with $r$ non-zero singular values on its main diagonal, where $r$ is the rank of the matrix $\mathbf{M}$ (38). Monitoring the rank deficiency of the single-value matrix $\mathbf{V}$ confirms the existence of multiple kinematic bifurcation points with their predicted locations at $\gamma_{12} = 90°$ and $180°$ and bifurcation paths (Fig. S3A-3B), which are in excellent agreement with the experimental observations (Fig. S3C).

Next, we assemble the reconfigurable cuboidal base units in a periodic way to form quasi-3D epithelial tissue-like metamorphic structures through in-plane tiling. It should note that periodic tiling of the units is non-trivial. Unlike traditional pin-joint tiling technique for 2D metamorphic mechanisms exhibiting only in-plane rotations (35), the base unit with 8R linkage possesses out-of-plane transformation. The challenge resides in achieving a tiling network with compatible kinematic motions to avoid structural frustrations (36), as well as inheriting the metamorphic gene in the base unit without sacrificing its intrinsic multiple kinematic modes and bifurcations.

To address the challenge, we design two fundamental building blocks by patterning the base unit (center) and its transformed state iii in Mode 3 into a " − " (Fig. 2A) and " + " (Fig. 2B) shape through merging the overlapped cubes uni-axially and bi-axially, respectively (see details on the hinge distribution in Fig. S4A). Such two building blocks will enable periodic assembly into compact and diluted tiling patterns as discussed later. Figure 2C-2D demonstrate their configuration evolutions through 3D-printed prototypes. To eliminate the potential torsion of their flexible hinges, we also fabricate the protypes of building blocks made of wooden cubes connected with metal hinges. We find that the kinematic transition in both prototypes are consistent with each other (Fig. S4B-S4D and Supplementary video 1).

Remarkably, both building blocks preserve all the three kinematic transition modes and bifurcations in the base unit without structural frustration, and consequently evolve into a library of distinct 3D transformable architectures (e.g., car, box, stair, building, airplane, and other complex 3D shapes in Fig. 2C-2D and Fig. S5-S6) by following each mode, i.e., bifurcated Mode 3 from 8R to 4R and to 2R/1R (Fig. 2C-i and Fig. 2D-i), and Mode 2 from 8R to 6R (Fig. 2C-iii and Fig. 2D-iii). However, transformations in Mode 1 diverge in the two building blocks (Fig. 2C-ii and Fig. 2D-ii). Referring to the extra 1R route in *yz*-plane (see the coordinate systems in Fig. 2C-i and 2D-ii) in the " + "-shaped building block, it exhibits one more type of kinematic transformation (i.e., Type 2 of Fig. 2D-ii-b and Fig. S6A-S6B) than the " − "-shaped one (Fig. 2C-ii and Fig. S5A). We note that during the transition, the additional hinges of overlapping-cuboids can act as a bridge to combine different transition modes for generating more types of configurations, e.g., a flying insect (top right of Fig. 2C-ii) combining 4R (state-iv or v of Mode 3 in Fig. 1D) and 1R linkage mechanism (middle left of Fig. 1B); and a hat combining 8R (state-i of Fig. 1D) and 1R linkage mechanism (middle right of Fig. 1B).

Next, based on the " + " and " − "-shaped building block, we create two periodic even-flat quasi-3D metamorphic materials with diluted (Fig. 3A, with voids) and compact (Fig. 3B, without voids) tiling networks, respectively. Both planar tiling patterns are composed of the base units (highlighted in grey color) and connectors (transformed state iii in Mode 3, highlighted in purple color). The design of connectors is key to ensuring compatible deformation between connected base units. To avoid geometric frustrations in the compact tiling, we implement two more extra pairs of hinges than the diluted tiling (see its opened state in right of Fig. 3B and hinge details in Fig. S7B). With the tiling deign, we note that both periodic networks disable transition Mode 2 of 6R linkage mechanism in the base unit due to the frustration with its contrary deployment features against the boundaries among the building blocks.

To inherit the kinematic transition Mode 3 of 8R linkage from the base unit, for the diluted pattern, the dihedral angles $\gamma_{b1}$ in the base unit and the connector $\gamma_{c2}$ should satisfy the compatibility condition of

$$\gamma_{b1} + \gamma_{c2} = 180º \qquad (1)$$

(see details of $\gamma_{b1}$ and $\gamma_{c2}$ in the right of Fig. 3A and Fig. S7A and S8A), while for the compact pattern, it needs to overcome the energy barrier (Fig. 3B and Fig. S9A-9B) caused by the twisted bridging hinges with incompatible transfigurations between the base units and connectors (38). We note that such a geometrical frustration can be readily avoided by sequentially deploying the structures via transition Mode 1 first and then Mode 3 (Fig. S9C-9D, Supplementary video 2 and 3). Thus, both tiling patterns can successfully inherit the primary kinematic features of diverse transition modes and bifurcations from the base unit.

Based on the unique tiling networks, we experimentally demonstrate their living-matter-like structural transition characteristics through metamorphic mechanisms in both diluted and compact models of prototypes (Fig. 3C-3F and Fig. S10-S11). To explore the effect of periodicity on the structural reconfigurations, for each tiling pattern, we fabricate two prototypes composed of different number of building blocks. Starting from an initial even-flat configuration, each tiling pattern can continuously evolve into varieties of exclusively 3D sophisticated architectures through the preserved kinematic transition Mode 1 (incomplete metamorphosis) and Mode 3 (complete metamorphosis with bifurcation).

We classify these evolved 3D configurations into two categories: one is metamorphic materials (MM) characterized of periodic structures such as MM-1, 2, 3, 4, and MM-5 in the diluted pattern as shown in Fig. 3C (2×2 units) and Fig. 3E (3×3 units), as well as MM-1', 2', and MM-3' in the compact pattern as shown in Fig. 3D (3×3 units) and Fig. 3F (5×5 units), where the same periodic 3D structures are observed in two protypes with different numbers of units, e.g., MM-1, 2 in Fig. 3C and 3E, and MM-1', 2' in Fig. 3D and 3F; the other is non-periodic 3D metamorphic architectures (MA) with specific topologies such as archway, truck, and buildings (Fig. 3C-3F), where a larger unit number leads to distinct and more complex architectures. We note that almost all these unique 3D metamorphic architectures (see more examples in Fig. S10-S11) are evolved from their metamorphic materials counterparts by rotating local structural segments. For example, truck-like shape in Fig. 3E can be achieved by first rotating boundary segments of MM-1 around $x$-axis followed by partially deploying $y$-directional structural segments (Fig. S12). Notably, the configurations of MM-3 (Fig. 3C-ii) and MM-2' (Fig. 3D-i, 3F-i) represent the bifurcation points in Mode 3, which can evolve along multiple branched kinematic paths to generate distinct metamorphic materials and 3D architectures. Such a transition from periodic metamorphic materials to significantly different non-periodic architectures is in analogy to complete metamorphosis in living systems, which involve the adaptive biological morphology evolution when confronting harsh environments interference (37).

To quantitatively characterize the inherence of kinematic features, we compare the structural kinematic mobilities during the transition of Mode 3 (insets of Fig. 3G) among the base unit and $2 \times 2$ units-based compact and diluted patterns. Fig. 3G shows that the kinematic bifurcations occur consistently with multi-DOFs at $\gamma_{12} = 0º$, 90º, and 180º, providing the evolutionary basis to create the metamorphic materials and architectures. It shows that the compact tiling pattern exhibits less mobilities than the diluted one due to the structural constraints by the introduced extra bridging-hinges.

To further quantify the complete metamorphosis process (via bifurcated transition Mode 3) in terms of structural mobility changes, we use the compact tiling prototype of 5×5 units as an example to evaluate its mobility evolution during its transition from configuration ① to configuration ⑦ (Fig. 3F). The results are shown in Fig. 3H (see details in (38)). It shows that the initial even-flat configuration ① has over 280 potential mobilities due to its bifurcation state. After reconfiguring to configuration ② and ③ through 8R linkage mechanism, the number of mobilities dramatically decreases to 1. Remarkably, the following configuration ④ possesses a peak number of over 16,000 mobilities arising from its kinematic bifurcations, which is over 57 times more than its initial state. As it further evolves, the number of mobility dramatically decreases to 3 at configuration ⑥ and finally reduces to 0 at configuration ⑦ which cannot reconfigure further.

The large number of mobilities at configuration ④ (i.e., MM-2' in Fig. 3f) endows a potential enormous space for structural reconfigurability. Given the respective number of independent $x$- and $y$-directional rotatable segments $N_x$ and $N_y$ in MM-2', we can calculate its combinatoric structural modes, which is equal to $2^{N_x+N_y} - 3$ ($N_x$, $N_y$ are integers; see Fig. 3i with $3 \leq N_x \leq 13$, $2 \leq N_y \leq 7$ and calculation details in (38)). For the model with a small number of 8×8 units, it can realize over $10^6$ structural modes with $N_x = 13$ and $N_y = 7$, offering a rich space to accomplish versatile intriguing structural evolutions.

The enormous combinatoric structural modes in periodic MM-2' (Fig. 3f) makes it a pluripotent platform for further programmable shape-morphing into 3D architected surfaces without or with intrinsic curvatures. To explicitly show the evolutionary shape-morphing process, we map the 3D MM-2' model onto a $v_n \times h_k$ 2D spin-frame network (Fig. 4a-iii), where its 3D unit cell composed of four rotatable wings around the rigid central part is shown on the top right (Fig. 4a-ii), and $v_n$ and $h_k$ are the number of rotatable parts along the vertical (highlighted in blue-dashed line in Fig. 4a-iii) and horizontal direction (highlighted in green-dashed line in Fig. 4a-iii), respectively. The symbols of blue-colored "↑, ↓" and green-colored "→, ←" represent the directional spin of rotatable side wings around the $(y^+, y^-)$ axis and $(x^+, x^-)$ axis, respectively. The free and independent rotation of four side wings in the 3D unit cells will generate four different combinatoric deployment mechanisms: 1-spin, 2-spin, 3-spin, and 4-spin, where $n$-spin represents rotating a number of $n$ wings that generates $C_4^n$ deployment motifs as illustrated in Fig. 4b (four sides wings are not shown in the deployment spin frame for clarity). Thus, through combinatoric deployment of different motifs among a number of unit cells in the initial spin-pattern, we can achieve a large number of programmable architectures with divergent topologies by following rational kinematic paths.

To demonstrate its programmable shape morphing of MM-2', starting from the initial spin-frame shown in Fig. 4c-i (only one quarter is shown for clarity considering its symmetry around both $x$- and $y$-axis), we realize a target architected pyramid shape (Fig. 4c-iii) through the forward design approach (*22*), see more details in Fig. S14. The detailed deployment spin-frame network is shown in Fig. 4c-ii including four combined spin motifs in the unit cell such as 1-spin (around $x^+$), two 2-spin (around $x^+$ and $x^-$; $x^+$ and $y^+$), and 3-spin (around $x^+$, $x^-$, and $y^+$) as illustrated in 3D in bottom right of Fig. 4c-ii. The kinematic path can be characterized with a shape function F(**R**), which takes the general form as:

$$F(\mathbf{R}) = \prod \mathbf{R}_{nx}(\theta_n)\mathbf{R}_{ky}(\theta_k) \qquad (2)$$

where **R** is the 3×3 right-bottom corner part of the transformation matrix **T**$_{ij}$ (see equation S2 in (38)), and $\Pi$ represents the multiplication operation of matrices. **R**$_{nx}$ and **R**$_{ky}$ represent the local spin-rotation matrix along *x*- and *y*-axis in each unit cell determined only by its related rotated angles 0°≤θ$_n$, θ$_k$ ≤90° (38). Since each product order represents one deployment path, given the independence of rotational spins, there are infinite kinematic paths to achieve the target shape by randomly arranging the product order of **R** in equation (2). Representatively, we decompose the spin motifs into two sequential deployment paths to generate the target pyramid shape: path 1 with *x*-spin first followed by *y*-spin (i.e., $F_{Path-1}(\mathbf{R}) = \sum_{n=1} \mathbf{R}_{nx} \sum_{k=1} \mathbf{R}_{ky}$) and path 2 with *y*-spin first followed by *x*-spin (i.e., $F_{Path-2}(\mathbf{R}) = \sum_{k=1} \mathbf{R}_{ky} \sum_{k=1} \mathbf{R}_{nx}$) with θ$_n$ = 90° and θ$_k$ = 90° (see Fig. S13 and the virtual model in Fig. S14). Furthermore, the same spin-frame network can also transform into an architected cuboid shape by rationally programming the spin rotation angles (Fig. 4c-iv).

Finally, starting from the pluripotent MM-2' platform, we propose an inverse design strategy to identify the prescribed kinematic paths to achieve the target more complex architected surfaces with intrinsic curvatures (22). To approximate a curved surface *S* with non-zero Gaussian curvature *K* such as the spherical shape (*K* > 0) in Fig. 4d, the inverse design procedure will follow 4 sequential steps: (1) discretizing the surface topology by two sets of parallel 2D planes (the dissecting planes are parallel to *xz* and *yz* plane defined in Fig. 4d-i) with identical intervals; (2) calculating two local curvatures ($k_x$, $k_y$) = ($\partial S/\partial x$, $\partial S/\partial y$) for each intersected point and mapping them to a periodic 2D square-shape networks (e.g., the discretized point *P* on a spherical surface is projected to point *P'* on *xy* plane, see details in Fig. 4d-i); (3) transforming each point on the mapping 2D network to its related local spin frame (including one base and two adjacent corner spins determined by the positive direction of ($k_x$, $k_y$), see details in Fig. S15a-ii) and generating the needed spin frame pattern (see details in Fig. S15a-i); (4) rotating the obtained spin frame via $n$-spin deployment mechanism (e.g., the zoomed 4-spin in Fig. 4d-ii) with rotated angles equal to those determined by two local curvatures ($k_x$, $k_y$) of each discretized node. By dissecting the spherical surface with 10 intervals along $x$ and $y$ axis, we demonstrate the success of the inverse design in obtaining the simulated approximately architected spherical shape from MM-2' shown in Fig. 4d-ii (see the related spin rotation angles for each discretized point in Table S1). In principle, by following the generalized inverse design strategy, starting from the same pluripotent MM-2' platform, we can achieve any curved surface topology with arbitrary intrinsic curvature. Representatively, we demonstrate the generation of more varieties of architected curved surfaces such as saddle shape with negative Gaussian curvature ($K < 0$) in a functional form of $S = x^2 - y^2$ ($0 \leq x, y \leq 1$) (Fig. 4e, spin rotation angles are listed in Table S2) and other ellipsoidal, wavy ribbon, and cylindrical shapes as shown in Fig. S15b. To validate the forward and inverse imitation capability of the MM-2', we experimentally assemble one MM-2' base pattern through multiple material 3D printed technique (see details in (38)), and based on which, demonstrate all the pyramid, sphere (partially) and saddle shapes (inserted pictures in Fig. 4c-iii, 4d-ii and 4e).

Extending spatial mechanisms, especially 3D base units with kinematic bifurcations, to origami-inspired architected materials design enables artificial materials with living-matter-like features equipped with metamorphic evolutionary capability to achieve one-to-$n$ various structure topologies. The proposed 3D spatial looped mechanism frees the constrained limited structural mobilities in conventional origami designs of planar thin sheets and thick-panel origami by spherical mechanism (*26, 27*). For practical applications of the designs, control and actuation of deployment will be important considering their rich mobilities. Exclusive linear kinematic paths for target shapes (e.g., the linear path in Fig. S13) will be needed for easy control of folding with motion actuators such as motors (*38*) or other stimuli-responsive actuators with only one DOF (*7*). This scale-independent mechanism design will find potential broad applications in designing reconfigurable architected metamaterials with tunable properties, evolutionary transformer robots adapting to changing environments, reconfigurable modular buildings, and deployable deep space infrastructure for rehabilitation.

**Acknowledgments:** The authors thanks Sid Collins for artistic inspiration. **Funding:** The authors acknowledge the funding support from National Science Foundation under award number CMMI CAREER-2005374. **Author contributions:** Y. L. and J. Y. proposed and designed the research. Y. L. designed and fabricated the prototypes and performed theoretical analysis of the system. J. Y. supervised the research. Y. L. and J. Y. wrote the manuscript. **Competing interests:** The authors declare no competing interests. **Data and materials availability:** All data is available in the main text or the supplementary materials.

**SUPPLEMENTARY MATERIALS**

Materials and Methods
Fig. S1 to S16
Table S1 to S2
Movie 1 to 3
Reference Notes

**Figures and Captions**

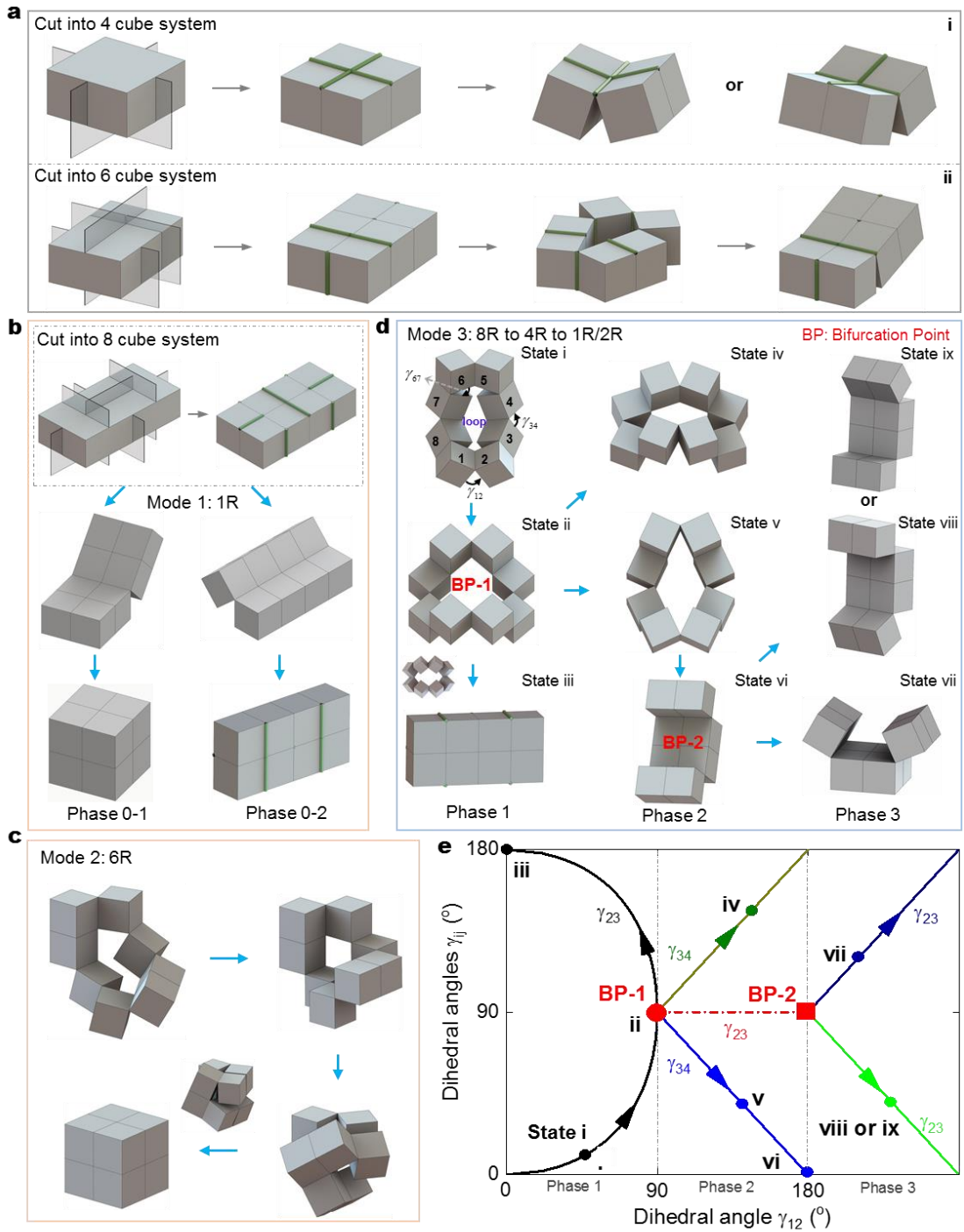

**Fig. 1. Systemic 3D modular kirigami designs by introducing cutting planes directly on 3D objects and their kinematic transitions.** (**a-b**) Schematics of the systemic 3D modular kirigami designs from 4 cubes to 8 cubes by introducing cutting planes directly on 3D objects: (**a**) two

selected 3D modular kirigami cutting with 4 and 6 cubes and their kinematic transitions; (**b**), the selected 3D modular kirigami cutting with 8 cubes, and its non-looped kinematic transition Mode 1 by rotating one pair(s) of line hinge with two transition paths without bifurcation. (**c**) Looped kinematic transition Mode 2 (6R) under geometric constraints of dihedral angles of $\gamma_{23} = \gamma_{67} = 0$ defined in (**c**) without bifurcation. (**d**) Looped kinematic transition Mode 3 (8R) with multiple bifurcation points (state ii and state vi) accompanied by multiple branched kinematic paths It bifurcates from 8R-linkage in Phase 1 (left column), to 4R-linkage in Phase 2 (middle column), and to 1R/2R-linkage in Phase 3. (**e**) Theoretically predicted evolution of dihedral angle changes during the transition Mode 3 shown in (**d**) through kinematic analysis.

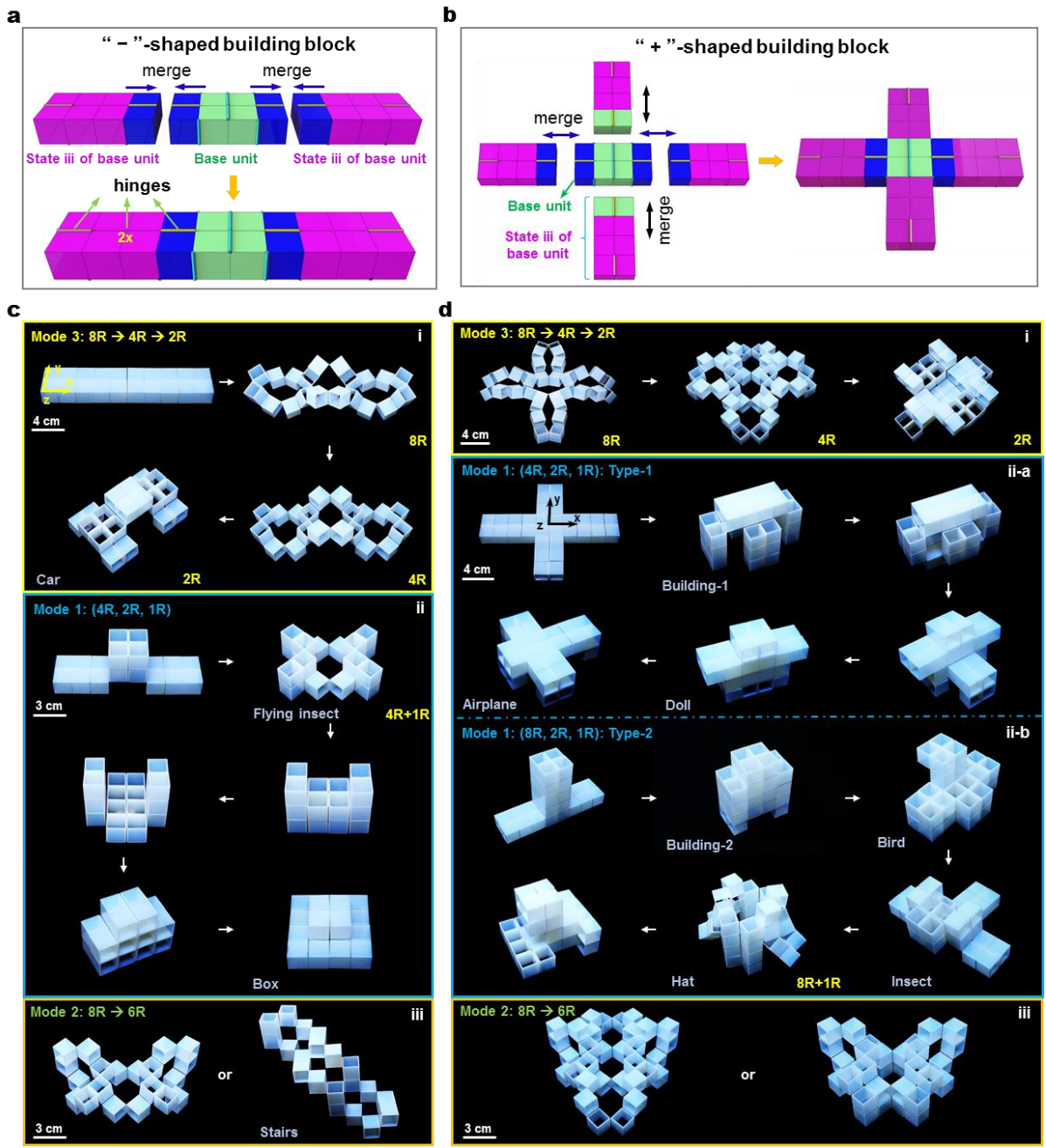

**Fig. 2. '−' and '+' shaped building blocks and their evolved morphologies.** (**a, b**) Schematic design details of the '−' and '+' shape building blocks by merging units uni-axially and bi-axially, respectively. (**c, d**) Prototypes of evolved structures of the two building blocks by following the three basic kinematic transition modes. **c**–*i* and **d**–*i* are based on kinematic transition Mode 3 with kinematic bifurcation, i.e. link number changes from 8R to 4R, finally to 2R/1R; **c**–*ii* and **d**–*ii*

correspond to kinematic transition Mode 1 and the evolved configurations are the combinatorics of base unit under reconfiguration with 2R and 1R linkage system, or with the 4R linkage system generated in Fig. 1**c**-*ii*; **c**–*iii* and **d**–*iii* relative to the kinematic transition Mode 3 given in Fig. 1**c** with linkage number remaining as 6R.

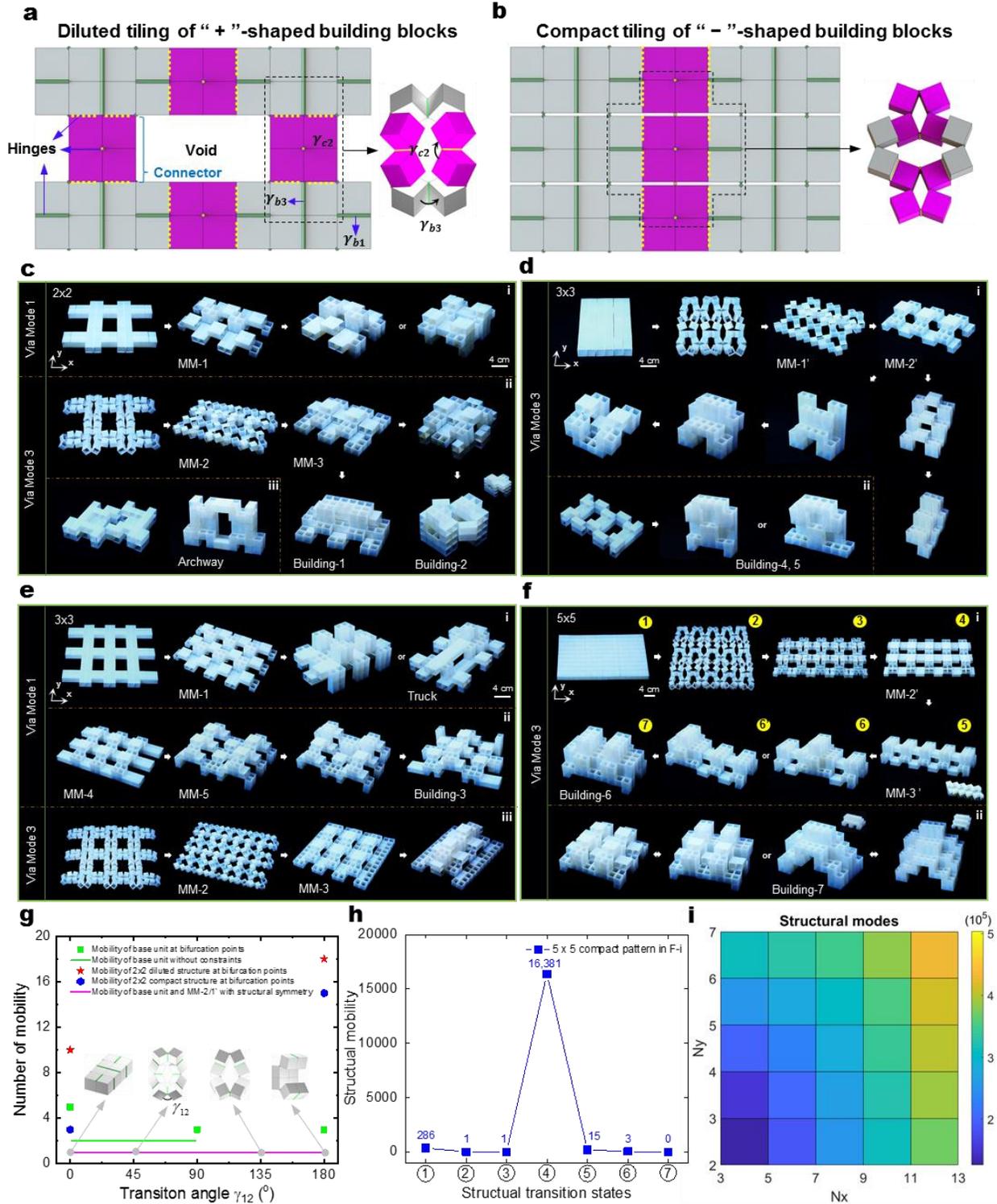

**Fig. 3. Metamorphic materials through periodic tiling of building blocks in diluted and compact tiling networks**. (**a**, **b**) Schematic design of the planar building blocks arrays and their connection details. (**c**, **d**, **e**, **f**) Prototypes of evolved 3D periodic materials and non-periodic

architectures and topologies from the two constructed periodic even flat tiling networks by following different kinematic modes. (**c, e**) Diluted patterns with $2 \times 2$ and $3 \times 3$ units. (**d, f**) Compact patterns with $3 \times 3$ and $5 \times 5$ units. (**g**) Comparison of kinematic mobilities between the base unit with/without structural symmetry constraint, and the $2 \times 2$ diluted and compact networks. (**h**) Evolution of the number of structural mobilities in $5 \times 5$ compact pattern of (**f**) during its complete metamorphosis reconfiguring from configuration ① to configuration ⑦. (**i**) Combinatoric structural modes of evolved MM-2' as shown in **d**-*i* and **f**-*i*.

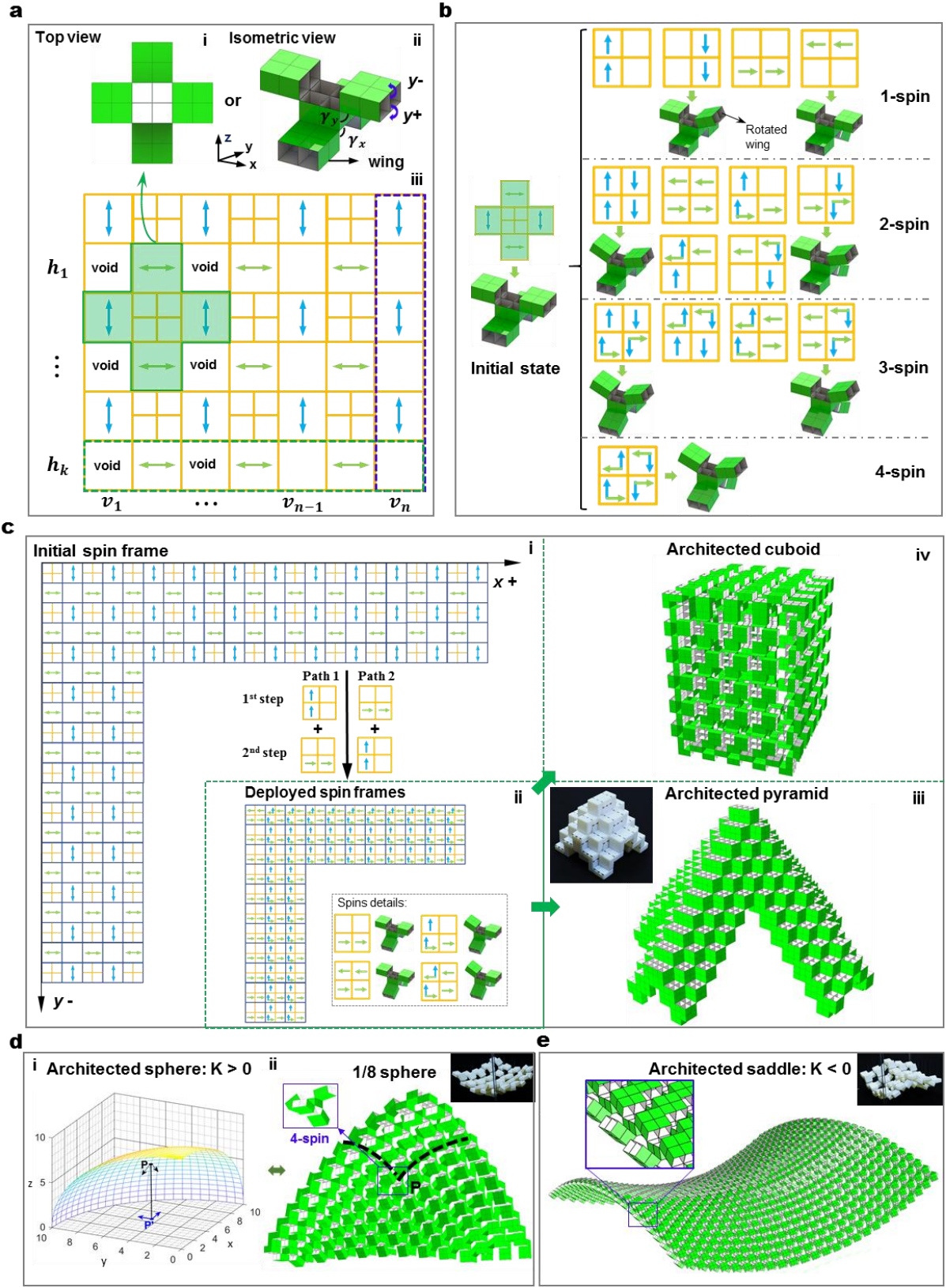

**Fig. 4 Forward and inverse designs for architected surfaces morphing from the pluripotent MM-2'platform.** (**a**) Mapped $v_n \times h_k$ 2D spin-frame network of the MM-2' platform. On the top: (i) and (ii) are the top view and isometric view of its virtual model of the periodic unit cell, respectively. (**b**) Four different rotated spin frames (1-spin, 2-spin, 3-spin, and 4-spin) by combinatorically rotating either 1, 2, 3, or 4 wings with schematically demonstrated virtual 3D models Selected virtual spin models. (**c**) Forward method to generate architected pyramid (iii with 3D-printed prototypes) and cuboidal (iv) surface topologies from the same initial (i) and deployed (ii) spin frame through two different sequential kinematic paths. (**d**) Inverse design to imitate 1/8 architected spherical surface topology with Gaussian curvature $K > 0$ from the pluripotent MM-**2'** platform whose spin frame is given in Fig. S15a. Left: discretization process of 1/8 spherical shape; Right: imitation of the discretized 1/8 spherical shape with 3D-printed samples. (**e**) The inverse imitated architected saddle surface (see also the inserted 3D-printed samples) with Gaussian curvature $K < 0$ based on the MM-2' platform.